\documentclass[12pt]{iopart}
\usepackage{graphicx}% Include figure files

\begin{document}

\title[]{Transport Theoretical Description of Collisional Energy Loss in Infinite Quark-Gluon Matter}

\author{Ghi R. Shin$^{1,2}$, Steffen A. Bass$^{2}$ and Berndt M\"uller$^{2}$}

\address{ $^{1}$Department of Physics, Andong National University,
Andong, South Korea}
\address{$^{2}$Department of Physics, Duke University, Durham, NC 27705, USA}

\date{\today}

\ead{gshin@phy.duke.edu}

\begin{abstract}
We study the time evolution of a high-momentum gluon or quark propagating through an 
infinite, thermalized, partonic medium utilizing a Boltzmann equation approach. 
We calculate the collisional energy loss of the parton, study its temperature and
flavor dependence as well as the the momentum 
broadening incurred through multiple interactions. Our transport calculations agree well
with analytic calculations of collisional energy-loss where available, but offer the unique
opportunity to address the medium response as well in a consistent fashion.
\end{abstract}

\submitto{\JPG}

\maketitle
%\newpage
\section{Introduction}

% goal is full jet evolution
% simulation of realistic scenarios
% requires validation vs. analytical results & benchmarks
% mention effect on medium (transport vs. jet MC)
% mention that bulk has already been studied, here emphasis on hard probes
% mention TECHQM

The currently prevailing view on the structure of the matter produced in nuclear collisions at the Relativistic Heavy-Ion Collider (RHIC) is anchored by two experimental observations \cite{Arsene:2004fa,Adcox:2004mh,Back:2004je,Adams:2005dq,Muller:2006ee}: (1) The emission of hadrons with a transverse momentum $p_T$ of several GeV/c or more is strongly suppressed of  (jet-quenching), implying the presence of matter of with a very large color opacity, and (2) The anisotropic (``elliptic'') flow in non-central collisions is near the ideal hydrodynamic limit, requiring an early onset of the period during which the expansion is governed by fluid dynamics (earlier than 1 fm/c after the initial impact) as well as nearly ideal fluid properties with a viscosity-to-entropy density ratio $\eta/s \ll 1$. The matter created at RHIC has been thus called the strongly interacting Quark-Gluon Plasma (sQGP) \cite{Gyulassy:2004zy}.

The origin of the jet-quenching phenomenon can be understood as follows \cite{Majumder:2010qh}: during the early pre-equilibrium stage of the relativistic heavy-ion collision, scattering of partons causing the formation of deconfined quark-gluon matter often engenders large momentum transfers which leads to the formation of two back-to-back hard partons. These traverse the dense medium, losing energy and finally fragment into hadrons which are observed by the experiments. Within the framework of perturbative QCD, the process with largest  energy loss of a fast parton is gluon radiation induced by collisions with the quasi-thermal medium. Elastic collisions add to the energy loss and are thought to be the dominant process for heavy quarks traversing the deconfined medium.

Even though the phenomenon is being referred to as "jet-quenching", the overwhelming majority of computations of this effect  have focused on the leading particle of the jet and do not take the evolution of
the radiated quanta into account. A variety of schemes for quantitative calculations of the radiative energy loss have been developed \cite{Zakharov:1996fv,Baier:1996kr,Gyulassy:1999zd,Wiedemann:2000ez,Guo:2000nz,Arnold:2001ba}. However, quantitative comparisons of theoretical model predictions 
incorporating a realistic hydrodynamic medium evolution with data from relativistic heavy ion collisions at RHIC \cite{Nonaka:2006yn} have revealed significant remaining ambiguities in the value of the extracted transport coefficients of the sQGP \cite{Bass:2008rv}. Different leading particle energy loss schemes may thus imply different values for these transport coefficients \cite{Majumder:2010qh}. A systematic comparison of the assumptions underlying the various leading particle energy loss schemes and of their  numerical implementations is currently under way \cite{Horowitz:2009eb,Majumder:2010qh}.

Recently, with the advent of sophisticated experimental techniques for the reconstruction of full jets emitted from an ultra-relativistic heavy-ion collision \cite{Bruna:2009em,Lai:2009zq}, attention has shifted from leading particle energy-loss to the evolution of medium-modified jets. The study of the evolution of the entire jet in the medium is expected to lead to a better understanding of the dynamics of energy-deposition into medium and of the subsequent medium response, e.g. the possible formation of Mach-cones etc. \cite{CasalderreySolana:2004qm,Neufeld:2009ep,Qin:2009uh}. The current state of the art in for medium modified jets are Monte-Carlo generators \cite{Lokhtin:2005px,Zapp:2009ud,Armesto:2009fj,Auvinen:2009qm}, which calculate the medium-modified jet but do not take the medium response into account. The consistent treatment of the jet in medium as well as the medium response requires the application of transport theory, e.g. via Boltzmann equation based calculation, such as is done in the Parton Cascade Model (PCM) \cite{Geiger:1991nj}. Parton Cascades have already been applied to the time-evolution of ultra-relativistic heavy-ion collisions. However, for the most part the PCM calculations have focused on reaction dynamics \cite{molnar:2000jh,bass:2002fh},  thermalization \cite{Xu:2004mz}, electromagnetic probes \cite{bass:2002pm,renk:2005yg}  and bulk properties of the medium \cite{molnar:2004yh}, as well as to a far lesser extent on leading particle energy-loss \cite{Fochler:2008ts,Fochler:2010wn}. We would also like to note the recent progress in combining a Boltzmann equation based
particle transport approach with pQCD cross sections for the scattering of high momentum particles
with soft particle interactions mediated by a Yang-Mills field \cite{Schenke:2008gg}.

It is our long-term goal to advance the application of the PCM to the description medium modified jets and the respective response of the medium to a jet propagating through it and depositing energy in it.  The achievement of this goal requires the validation of the PCM against analytic test cases which can be reliably calculated for a simplified medium, e.g. a pure gluon plasma or a quark-gluon plasma in thermal and chemical equilibrium. In the present manuscript we take a first step in this direction. We calculate elastic energy loss in an infinite, homogeneous medium at fixed temperature within the PCM approach and compare our results to analytic calculations of the same quantity. In addition, we calculate the rate of momentum broadening of a hard parton propagating through the medium and compare the results of our analysis to analytic expressions for the transport coefficient $\hat q$.

%%%%%%%%%%%%%%%%%%%%%%%%%%%%%%%%%%%%%%%%%%%%%%%%%%%%%%

%\newpage
\section{Quark Gluon Plasma and Parton Cascade Simulations}

The medium in our calculations is an ideal Quark-Gluon-Plasma ,i.e. a gas of
$u, d$ and $s$ quarks and anti-quarks as well as gluons at fixed temperature $T$ 
in full thermal and chemical equilibrium. In addition, we also conduct studies for a one-component 
gluon plasma in thermal and chemical equilibrium. For our transport calculation we
define a box with periodic boundary conditions (to simulate infinite matter) and 
sample thermal quark and gluon distribution functions to generate an ensemble of
particles at a given temperature and zero chemical potential. We then insert
a hard probe, i.e. a high momentum parton, into the box and track its evolution through
the medium. The medium particles may interact with the probe as well as with each
other according to a Boltzmann Transport equation:

\begin{equation}
p^\mu \frac{\partial}{\partial x^\mu} F_i(x,\vec p) = {\cal C}_i[F]
\label{eq03}
\end{equation}
where the collision term ${\cal C}_i$ is a nonlinear functional 
of the phase-space distribution function. Although the collision
term, in principle, includes factors encoding the Bose-Einstein 
or Fermi-Dirac statistics of the partons, we neglect those effects
here.

The collision integrals have the form:
\begin{equation}
\label{ceq1}
{\cal C}_i[F] = \frac{(2 \pi)^4}{2 S_i E_i} \cdot
\int  \prod\limits_j {\rm d}\Gamma_j \, | {\cal M} |^2 
       \, \delta^4(P_{\rm in} - P_{\rm out}) \, 
          D(F_k(x, \vec p)) 
\end{equation}
with
\begin{equation}
D(F_k(x,\vec p)) \,=\, 
\prod\limits_{\rm out} F_k(x,\vec p) \, - \,
\prod\limits_{\rm in} F_k(x,\vec p) \quad
\end{equation}
and
\begin{equation}
\prod\limits_j {\rm d}\Gamma_j = \prod\limits_{{j \ne i} \atop {\rm  in,out}} 
        \frac{{\rm d}^3 p_j}{(2\pi)^3\,(2p^0_j)} 
\quad.   
\end{equation}
$S_i$ is a statistical factor defined as
$S_i \,=\, \prod\limits_{j \ne i} K_a^{\rm in}!\, K_a^{\rm out}!$
with $K_a^{\rm in,out}$ identical partons of species $a$ in the initial
or final state of the process, excluding the $i$th parton.

The matrix elements $| {\cal M} |^2$ account for the following 
processes:
\begin{equation}
\label{processes}
\begin{array}{lll}
\,g g \to g g \quad&\quad q q \to q q \quad&\quad	q g \to q g \\
\,q q' \to q q' \quad& \quad q \bar q \to q \bar q \quad& \quad	\bar q g \to \bar q g\\
(g g \to q \bar q) \quad& \,\,\,\,(q \bar q \to g g) \quad& \,\,\,\,(q \bar q \to q' \bar q') \\
\end{array}
\end{equation}
with $q$ and $q'$ denoting different quark flavors. The flavor changing processes
in parenthesis are optional and can be disabled to study the effect of jet flavor conversion.
The gluon radiation
processes, e.g. $g g \rightarrow g g g$ are not included in this study, but will be addressed
in a forthcoming publication. 
The amplitudes for the above processes have been calculated in refs. 
\cite{Cutler:1977qm,Combridge:1977dm} for massless quarks. The 
corresponding scattering cross sections are expressed in terms
of spin- and colour-averaged amplitudes $|{\cal M}|^2$:
\begin{equation}
\label{dsigmadt}
\left( \frac{{\rm d}\hat \sigma}
     {{\rm d} Q^2}\right)_{ab\to cd} \,=\, \frac{1}{16 \pi \hat s^2}
        \,\langle |{\cal M}|^2 \rangle
\end{equation}
For the transport calculation we also need the total cross section 
as a function of $\hat s$ which can be obtained from (\ref{dsigmadt}):
\begin{equation}
\label{sigmatot}
\hat \sigma_{ab}(\hat s) \,=\, 
\sum\limits_{c,d} \, \int\limits_{(p_T^{\rm min})^2}^{\hat s}
        \left( \frac{{\rm d}\hat \sigma }{{\rm d} Q^2}
        \right)_{ab\to cd} {\rm d} Q^2 \quad .
\end{equation}

Since our medium is in full thermal and chemical equilibrium, 
we can use the effective thermal mass of a gluon and a quark in the system to
regularize the cross sections \cite{biro:1993qt}:
\begin{eqnarray}
\mu_D^2 &=& \pi \alpha_s d_p \int {{d^3 p}\over{(2\pi)^3}}
{{C_2}\over{|\vec p|}} f_p(\vec p ; T),
\end{eqnarray}
where $d_p$ is the degeneracy factor of a parton p and 
$C_2$ is $N_c$ for gluons and $(N_c^2-1)/(2N_c)$ for quarks. Inserting
the thermal distribution yields a
 Debye mass of a gluon of $\mu_D = g T$ in a thermal gluon system at temperature $T$ and of 
$\mu_D = \sqrt{ (2 N_c + N_f)/6}  g T$ for quarks and gluons in a quark-gluon plasma.
For example, the dominant elastic cross sections thus are: 
\begin{eqnarray}
\label{diffxs}
{{d\sigma^{gg\rightarrow gg}}\over{dq_\perp^2 }} &=& 
{2\pi\alpha_s^2}  {9 \over 4} {1 \over {(q_\perp^2 + \mu_D^2 )^2}},
\label{diff_cs_gg_gg} \\
{{d\sigma^{gq\rightarrow gq}}\over{dq_\perp^2 }} &=& 
{2\pi\alpha_s^2} {1 \over {(q_\perp^2 + \mu_D^2 )^2}},
\label{diff_cs_gq_gq} \\
{{d\sigma^{qq\rightarrow qq}}\over{dq_\perp^2 }} &=& 
{2\pi\alpha_s^2}  {4 \over 9} {1 \over {(q_\perp^2 + \mu_D^2 )^2}},
\label{diff_cs_qq_qq}
\end{eqnarray}
where the first order of Feynman diagrams have been included.

For our studies we use two distinct implementations of the Parton Cascade Model, the
the Andong parton cascade code \cite{shin:2002fg}, and the VNI/BMS code \cite{bass:2002fh}. Both
codes have been modified to contain the same cross sections so that we can verify
the outcome of our calculations through cross-comparisons between the two PCM 
implementations. For the sake of simplicity we keep the coupling constant fixed at a 
value of $\alpha_s = 0.3$.

To obtain the temporal evolution of a high momentum parton 
propagating through a gluon or a quark-gluon plasma,
a high energy gluon or quark of initial energy $E_0$ is injected into or thermal (quark-)gluon-plasma
box at the center of it's xy-plane with ${\vec p}=(0,0,p_z=\sqrt{E_i^2})$. Following each individual
scattering event involving our hard probe, we record the energies and momenta of the outgoing
partons. The parton with the larger momentum in the final state is considered to be our
continuing probe particle, due to the dominance of small angle scattering in the 
implemented cross sections.
A simple estimate on the number of interactions the probe will undergo per unit time or
length can be obtained via its mean free path in the medium:
\begin{eqnarray}
\lambda_s &=& {{1}\over {\sigma_T \rho}},
\end{eqnarray}
where $\sigma_T$ is the total thermal cross section and $\rho$ the density of a medium.
A gluon's mean free path in a gluon plasma is $1.1$~fm at $T=300$~MeV and $0.76$~fm at
$T=400$~MeV, and in the quark-gluon plasma  $0.83$~fm and $0.59$~fm, respectively.
The quark mean free paths are about $9/4$ times longer than those of gluon ones at same temperature.

\section{Results and Discussion}

\begin{figure}[t]
\begin{center}
\includegraphics[width=.95\textwidth]{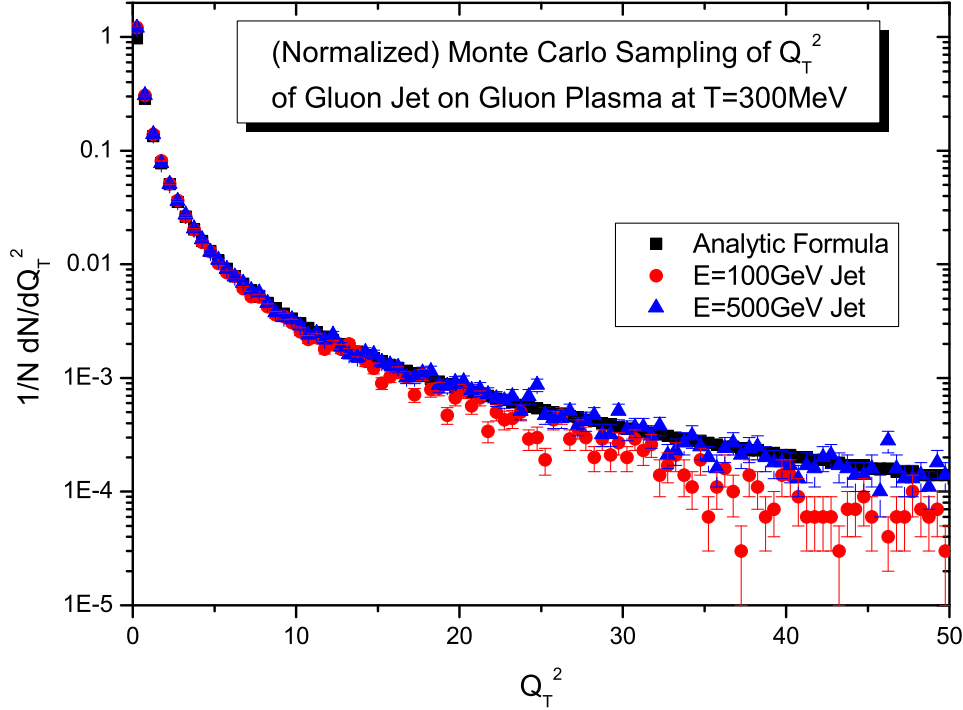}
\caption{The Monte Carlo sampling of transverse momentum transfer
at each scattering. The jet only is allowed to have a collision with medium particles.}
\label{fig1}
\end{center}
\end{figure}

In our work, we shall focus solely on the hard parton propagating through the 
medium and analyze its energy loss and momentum broadening as a function of
time and distance traveled. We would like to point out that our calculation 
includes the full information on the medium-response
as well, however, a detailed analysis on its characteristics it outside the
scope of the present work and will be discussed in a forthcoming publication.

Let us start by comparing the distribution of transverse momentum transfers experienced by  
gluons with $E_0= 100$~GeV and $E_0=500$~GeV, respectively, 
while propagating through a gluon plasma at $T=300$~MeV to the analytic expression given
by the differential cross section (\ref{diffxs}).

Figure \ref{fig1} shows that we find excellent agreement between our transport calculation
and the analytic expression for $E_0= 500$~GeV. At the lower incident probe energy, 
$E_0=100$~GeV we note a suppression in the high momentum transfer tail of the distribution 
which we can attribute to the effects of phase space and energy conservation,
the gluon loses a substantial amount of energy while traversing
a medium of 50 fm depth and thus for most of the time does not carry sufficient momentum to
scatter with high momentum transfers while the analytic
formula continuously assumes that the CM energy between the colliding particles 
is much larger than the Debye mass.

\begin{figure}[t]
\begin{center}
\includegraphics[width=0.49\textwidth]{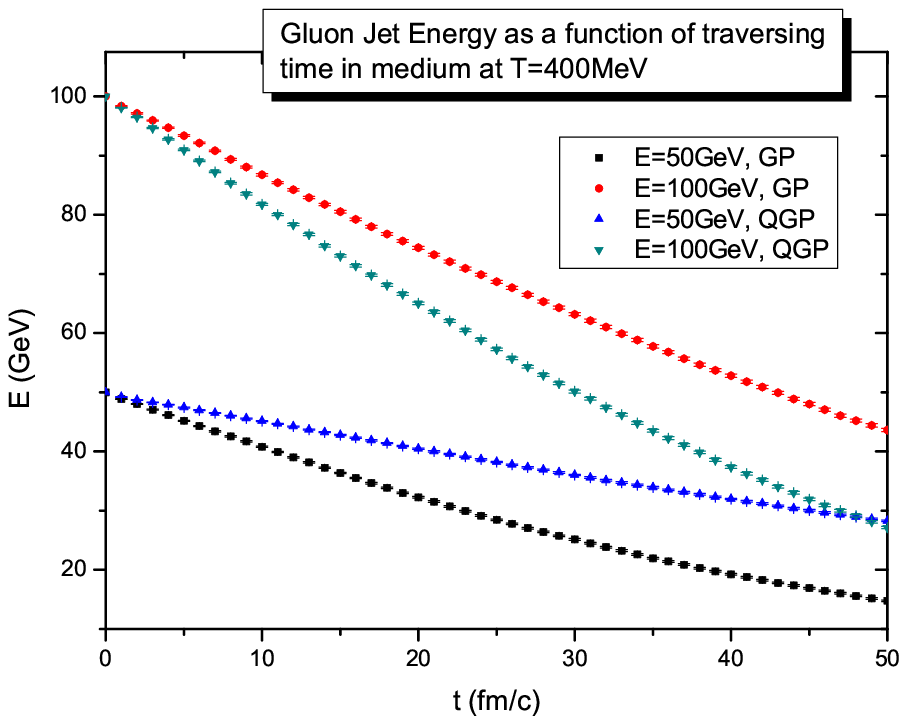}\hfill
\includegraphics[width=0.49\textwidth]{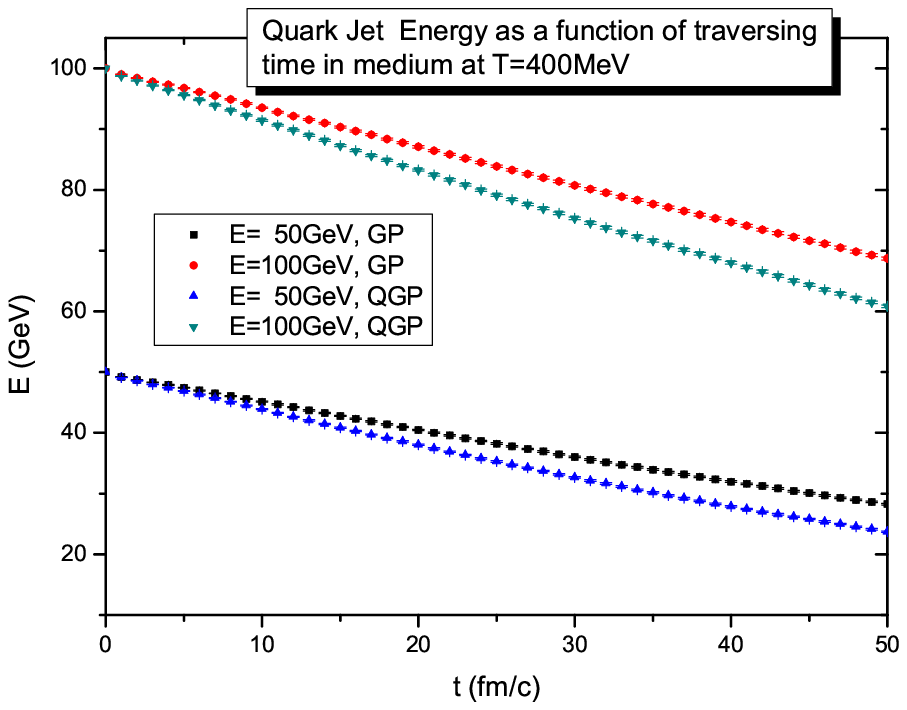}
\caption{Left: the gluon jet energy as a function of traveling time in the medium at $T=400MeV$: 
The travelling distance is proportional to the time. GP is a gluon plasma and QGP a quark-gluon
plasma and the simulation does not include $gg \rightarrow ggg$ process.
Right: the quark jet energy as a function of time: The penetration length
is proportional to the time. Plasma temperature $T=400MeV$.}
    \label{fig2}
\end{center}
\end{figure}

We now focus on the elastic energy loss of a high-momentum parton in medium. 
The left frame of Fig. \ref{fig2} shows the energy as a function of time for a gluon
with initial energy of 100~GeV (or 50~GeV, respectively), propagating through a 
quark-gluon plasma (QGP) or gluon plasma (GP) at a temperature of T=400~MeV. 
The right frame repeats the calculation for a light quark instead of a gluon.
The calculation, which includes only elastic processes (i.e. the flavor exchange reaction
channels have been disabled in order to unambiguously study the flavor dependence) clearly
shows the anticipated linear decrease of jet energy with time. One should note, that analytic
calculations usually study the jet energy as a function of distance traveled -- in our case
we substitute time for distance in order to have a quantity that is not affected by the 
periodic boundary conditions for the system in our calculations. We have verified the linear
relationship between distance traveled and elapsed time with a slope near unity.

The two frames clearly show the difference between a GP and a QGP in terms of energy-loss
at the same temperature. This difference is due to the significantly higher parton density
in a QGP compared with a GP for the same temperature, resulting in a larger number of scattering
partners for the hard probe to interact with. The difference is less pronounced for the
quark probe than the gluon probe, due to the difference in their interaction cross sections.
We also observe that the linear decrease of energy as a function of time tapers off  
in the long time limit as the probe energy approaches the thermal regime. While at the probe
energies studied here (between 50~GeV and 400~GeV) this occurs at times not relevant
in the context of an ultra-relativistic heavy-ion collision, for smaller probe energies
frequently seen at RHIC, this may be a significant effect.

\begin{figure}[t]
\begin{center}
\includegraphics[width=0.48\textwidth]{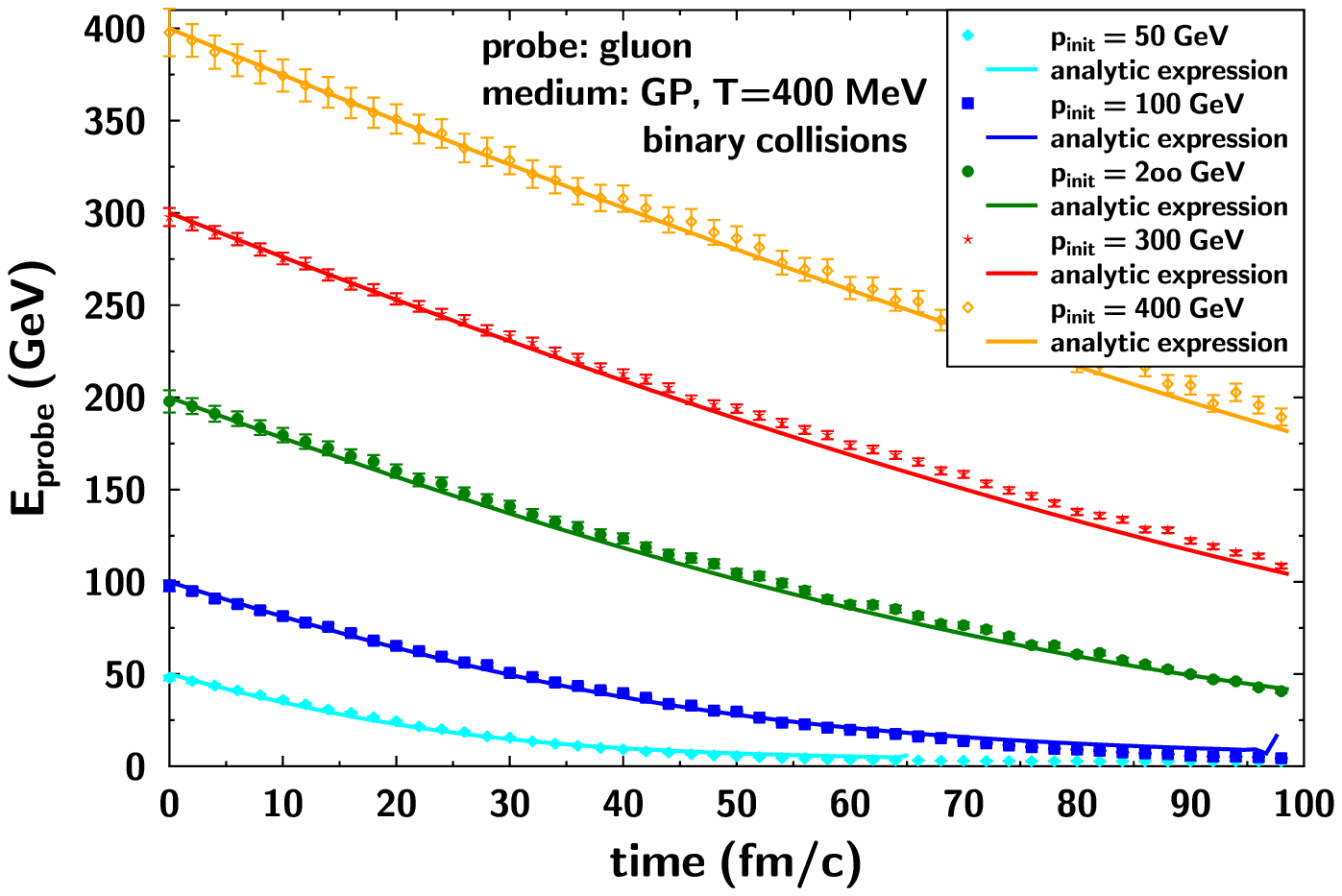}\hfill
\includegraphics[width=0.48\textwidth]{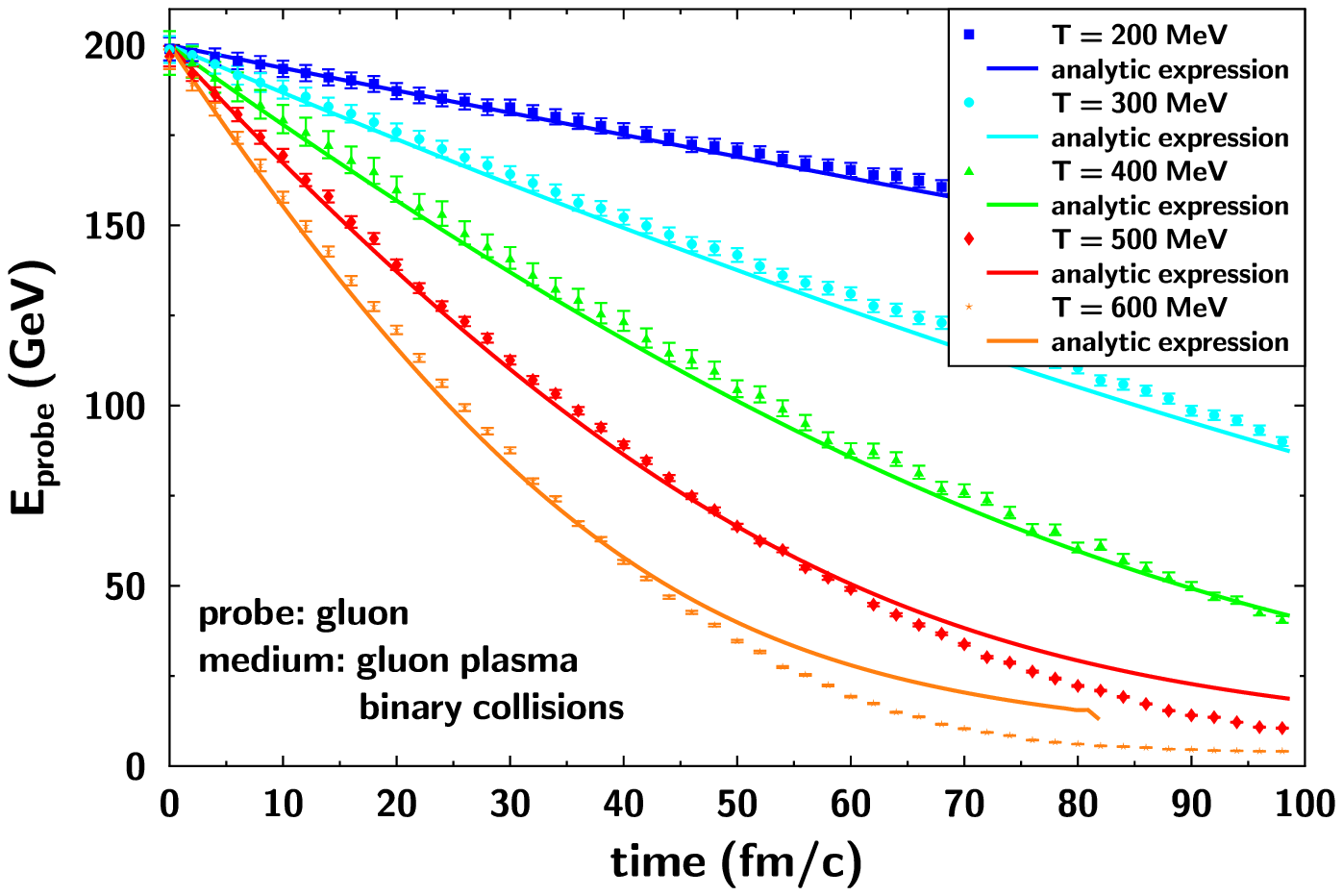}
\caption{Left: gluon energy as a function of time for different initial values of the energy in a 
gluon plasma. Right: the same for a fixed initial energy and various medium temperatures. The solid
lines represent an analytical calculation (see text for details).}
    \label{fig3}
\end{center}
\end{figure}

Fig.~\ref{fig3} shows the energy and temperature dependence of the elastic energy loss of
a hard gluon in a gluon plasma. The calculations (symbols) are compared to an analytical 
calculation \cite{bjorken:1982tu,thoma:1992kq}:
\begin{equation}
- {{dE}\over{dt}} = \int {{d^3 k}\over{(2\pi)^3}} F_g(\vec k;T)
\int dq_\perp^2 (1-\cos\theta) \nu {{d\sigma}\over{dq_\perp^2}}
\label{e_loss_1}
\end{equation}
where $\nu = E-E'$ is the energy difference between before and after collision.
Utilizing the characteristics of our medium and the cross sections in our calculation, this 
equation can be discretized to a form which lends itself to a comparison
with our calculation:
\begin{equation}
E_p(z) = E_p(0) - z \frac{\alpha_s C_2 \mu_D^2}{2} \ln\left[
\frac{\sqrt{E_p(z) T}}{\mu_D} \right]
\label{dedx}
\end{equation}
Using an iterative procedure we can calculate $E_p(z)$ for the initial gluon energies
and temperatures used in \mbox{Fig. ~\ref{fig3}} and compare them to our calculation. The
agreement between the analytical calculation and the PCM is remarkable and serves as 
validation of the PCM framework. The validation of the PCM calculation
in a well-controlled infinite matter calculation in full equilibrium at fixed temperature
is of significant importance, since the PCM can easily be used to study realistic dynamic systems
far off equilibrium, e.g. an ultra-relativistic heavy-ion collision, which can only be 
poorly described in the framework of (semi-)analytic calculations.

\begin{figure}[t]
\begin{center}
\includegraphics[width=.48\textwidth]{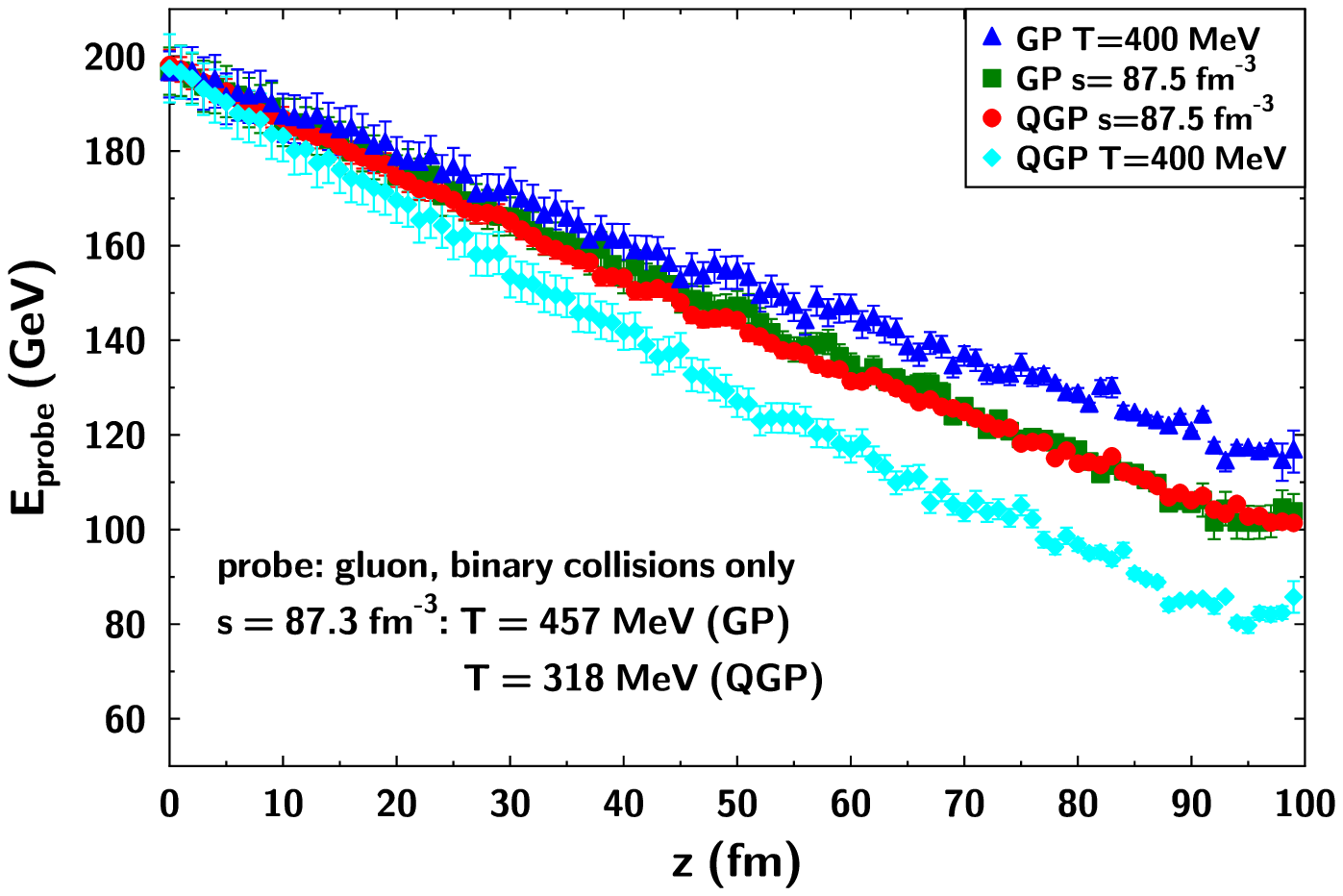}\hfill
\includegraphics[width=0.49\textwidth]{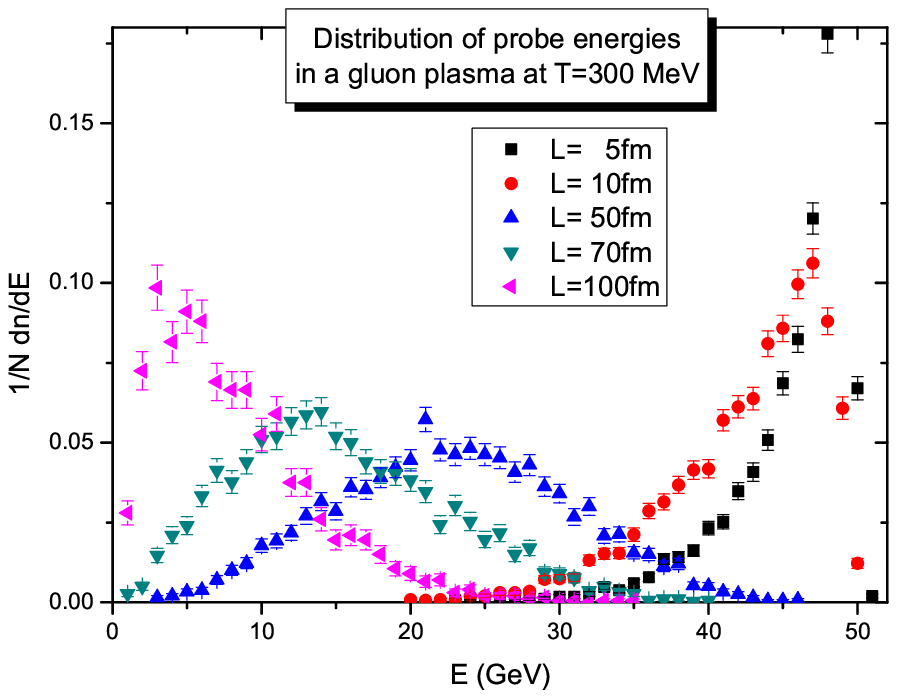}
\caption{Left: Energy as a function of distance for a gluon propagating through
a GP and a QGP at the same temperature and at equivalent entropy-densities.
A scaling of the energy-loss with the entropy-density is observed. Right: distribution
of probe energies after traversing the medium for 5, 10, 50, 70 and 100 fm at a temperature
of 300~MeV.}
\label{fig4}
\end{center}
\end{figure}

In \mbox{Fig. ~\ref{fig2}} we observe a difference in the energy loss a parton suffers
when propagating through a GP or a QGP. We understand this difference to be due
to the different overall particle densities associated with a GP and a QGP at the same
temperature. In order to validate this point, we initialized a GP and a QGP at an
identical entropy density of $s = 87.5$~fm$^{-3}$, corresponding to a temperature
of $T=457$~MeV for a GP and $T=318$~MeV for a QGP. The results of this calculation 
(also in comparison to a GP and a QGP at a temperature of $T=400$~MeV)
are shown in \mbox{Fig.~\ref{fig4}}: for the same entropy density, both the GP and the QGP
inflict a nearly identical amount of elastic energy loss to the hard probe. This scaling
suggests that the entropy density, rather than the temperature, may be a robust quantity 
for the characterization of a thermal QCD medium by energy-loss measurements. We note
that it is by no means clear whether the deconfined medium in the early stages 
of a heavy-ion reaction more closely resembles a GP (or a liberated ``glasma'' 
\cite{Lappi:2006fp}) or a QGP in full chemical equilibrium \cite{biro:1993qt}.
Most likely the chemical composition of the deconfined medium created in a ultra-relativistic
heavy-ion collision changes significantly as a function of time, whereas its entropy
density (after the initial equilibration) does not vary as much.

The right frame of \mbox{Fig.~\ref{fig4}} shows the distribution of probe energies for a gluon
with initial energy of 50~GeV passing
through a gluon plasma at temperature $T=300$~MeV after 5, 10, 50, 70 and 100~fm, respectively.
The gluon distribution at $t=0$ is a $\delta$-function at $E=E_0$. 
The Figure shows several interesting features. We find that elastic energy loss acts 
as a diffusion processes
in momentum space, with the width of the distribution being a function of the traveling distance.
Likewise the peak of the distribution shifts as a function of traveling distance.
Even after traveling significant distances, about an order of magnitude larger than possible in a
relativistic heavy-ion collision, the gluon has not thermalized with the medium, 
in which a medium particle has an average energy of about 0.9 GeV at T=300~MeV.

%\clearpage
Another important quantity for the characterization of the hot and dense QCD medium is
the transport coefficient $\hat q$, which is defined as \cite{majumder:2007zh}: 
\begin{eqnarray}
\hat{q}_R &=& \rho(T) \int dq_\perp^2 q^2 {{d\sigma}\over{dq_\perp^2}}
\label{q_hat}
\end{eqnarray}
with the squared momentum transfer  $q^2 = -t$. Generally, $\hat q$ can be interpreted
as the amount of squared transverse momentum per unit length a probe accumulates
as it propagates through the QCD medium. This interpretation lends itself to a definition
of $\hat q$ suitable for extraction from microscopic transport calculations:
\begin{equation}
\hat q \,=\, \frac{1}{l_z} \sum \limits_{i=1}^{N_{coll}} \left(\Delta p_{\perp,i}\right)^2
\end{equation}
One should note that $\hat q$ in general should depend on the distance $l_z$ the probe has
traveled through the medium, since the probe loses energy in the process, and thus the
average momentum transfers in individual interactions of the probe with the medium may
vary as a function of that distance.

\begin{figure}
\begin{center}
\includegraphics[width=0.9\textwidth]{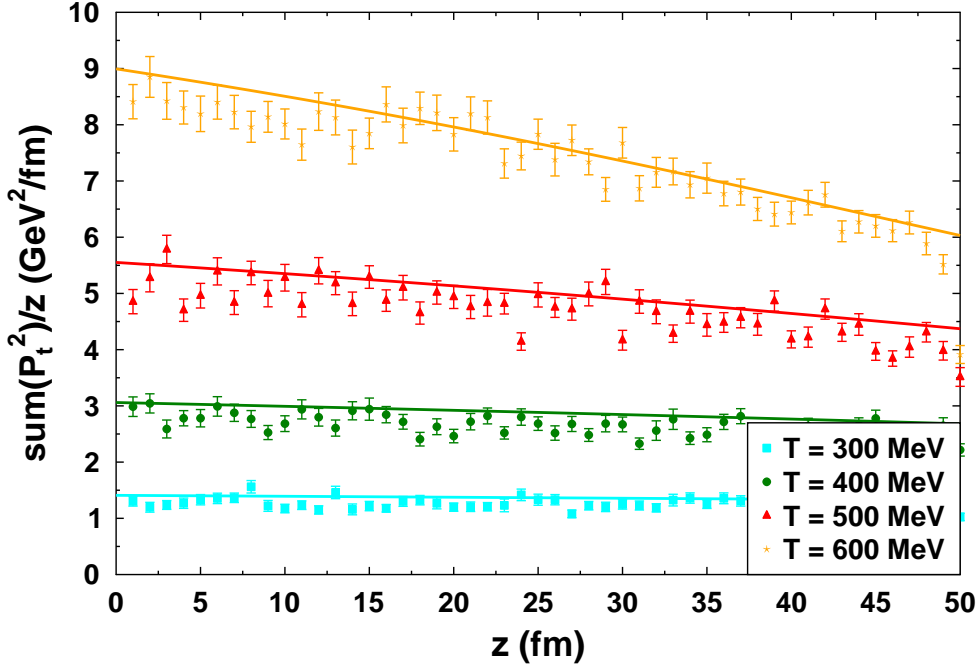}\hfill
\caption{Sum of squared transverse momentum transfers per unit length as a function
of distance traveled for a 50 GeV gluon in a gluon plasma at different temperatures.
This quantity corresponds to the transport coefficient $\hat q$.}
    \label{fig5}
    \end{center}
\end{figure}

Figure~\ref{fig5} shows $\hat q$ for a gluon plasma as a function of distance
traveled through the medium for several temperatures. The initial values of $\hat q$ depend
strongly on the temperature, ranging from approximately 1.5 GeV$^2$/fm at $T=300$~MeV up to 
about 9 GeV$^2$/fm at 600~MeV temperature. For $T=400$~MeV we find good agreement
with recent calculations by \cite{Schenke:2008gg} using a Boltzmann transport for hard
binary collisions combined with soft interactions mediated by a collective Yang-Mills field
and by \cite{Fochler:2010wn} with the BAMPS parton cascade model in its binary
scattering mode. 
At leading-log order and in the weak coupling limit,
using the same matrix element as in our PCM calculation,
the following analytic expression has been derived for $\hat q$ \cite{Arnold:2009ik}:
\begin{equation}
\hat q (\Lambda) \, \approx \, \alpha_s\, T\, m_D^2 \ln\left(\frac{T^2}{m_D^2}\right) + 4 \pi \, \alpha_s^2 \, {\cal{N}}
\ln\left(\frac{\Lambda^2}{T^2}\right)
\label{qhat}
\end{equation}
if the cut-off $\Lambda \geq T$. Comparing expression~(\ref{qhat}) to expression~(\ref{dedx}) and
following the derivation of~(\ref{qhat}) we find that the cut-off $\Lambda^2 \sim E_p T$. Replacing
$\Lambda^2$ by $a E_p(z) T$ (with $a$ being a proportionality constant to be determined) 
in expression~(\ref{qhat}) provides a good fit to our results for a choice of $a=0.1$.

\section{Conclusions}
We have studied elastic energy loss of high energy parton in an infinite, homogeneous, thermal medium  within the PCM approach and have compared our results to analytic calculations of the same quantity. In addition, we have calculated the rate of momentum broadening of a hard parton propagating through the medium and have compared the results of our analysis to an analytic expression for the transport coefficient $\hat q$. We find good agreement between the PCM calculations and the analytic
expressions (within the approximations used in both cases), giving us significant confidence that
our transport approach provides a reliable description of a gas of quarks and gluons at temperatures
above $\approx 2 T_C$ in the weak coupling limit. We expect that the results presented in this
manuscript can be used as a benchmark by other parton-based microscopic transport calculations.
The validation of the PCM against the analytic test cases presented in this manuscript now allows
us to  advance the application of the PCM to the description of medium modified jets in relativistic 
heavy-ion collisions and the response of the medium to a hard parton propagating through it.

\section{Acknowledgements} Ghi R. Shin was supported by a grant from Andong National University (2008)
and thanks the members of the Physics Department at Duke University for their warm hospitality 
during his sabbatical visit. S.A.B. and B.M. were supported by the DOE under grant DE-FG02-05ER41367.
The calculations were performed using resources provided by the Open Science Grid, 
which is supported by the National Science Foundation and the U.S. Department of 
Energy's Office of Science.

%\clearpage

\section*{Refences}
\bibliographystyle{iopart-num}
\bibliography{/Users/bass/Publications/SABrefs}

\end{document}